\documentclass[aps,amssymb,twocolumn]{revtex4}%,showpacs
\usepackage{amssymb}
\usepackage{graphicx}
\usepackage{color}
\usepackage{textcomp}
\usepackage{ulem}
\begin{document}

\title{ Pion superfluid phase transition under external magnetic field including inverse magnetic catalysis effect}
\author{Shijun Mao }
 \email{maoshijun@mail.xjtu.edu.cn}
\author{Yvming Tian}
\affiliation{School of Physics, Xi'an Jiaotong University, Xi'an, Shaanxi 710049, China}

\begin{abstract}
Pion superfluid phase transition under external magnetic field including the inverse magnetic catalysis (IMC) effect is investigated by the Pauli-Villars regularized NJL model. Based on the Goldstone's theorem, we apply the massless Goldstone boson ($\pi^+$ meson) to determine the onset of pion superfluid phase. The inverse magnetic catalysis effect is introduced by the magnetic field dependent coupling $G(eB)$, which is a decreasing function of magnetic field. At fixed temperature and baryon chemical potential, the critical isospin chemical potential for pion superfluid phase transition including IMC effect increases as the magnetic field grows, which is similar as the case without IMC effect. This demonstrates that magnetic field disfavors the pion superfluid phase when considering or ignoring IMC effect. The critical isospin chemical potential at fixed magnetic field, temperature and baryon chemical potential is shifted to higher value by the IMC effect. Since it is more difficult to form pion superfluid with weaker coupling.
\end{abstract}

\date{\today}
%\pacs{12.38.-t, 25.75.Nq, 14.80.Mz}%11.30.Rd, 14.40.-n, 21.65.Qr
\maketitle

\section{Introduction}
Recently, the magnetic field effect on QCD phase structure attracts much attention~\cite{review0,review1,review2,review3,review4,review5} due to its close relation to high energy nuclear collisions and compact stars. For instance, the LQCD simulations performed with physical pion mass observe the inverse magnetic catalysis (IMC) phenomenon~\cite{lattice1,lattice11,lattice2,lattice3,lattice4,lattice5,lattice6,lattice7}. Namely,
the pseudo-critical temperature $T_{pc}$ of chiral symmetry restoration and the quark mass near $T_{pc}$ drop down with increasing magnetic field. On analytical side, many scenarios are proposed to understand this inverse magnetic catalysis phenomenon, but the physical mechanism is not clear~\cite{fukushima,mao,kamikado,bf1,bf12,bf13,bf2,bf3,bf4,bf5,bf51,bf52,bf8,bf9,bf11,db1,db2,db3,db5,db6,pnjl1,pnjl2,pnjl3,pnjl4,pqm,ferr1,ferr2,mhuang,meijie,ammmao}.

QCD phase structure at finite isospin chemical potential contains the chiral symmetry restoration and pion superfluid phase transitions. With vanishing external magnetic field and temperature, when the isospin chemical potential is higher than the pion mass in vacuum, the $u$ quark and $\bar d$ quark form coherent pairs and condensate. The system enters the pion superfluid phase and the charged pion becomes massless as the corresponding Goldstone mode~\cite{lqcd1,lqcd2,lqcd3,model1,model2,model3,model4,model5,model6,model7,model8,model9,model10,model11,model12,model13,
model14,model15,model16,model17,model18,model19,model20,model21,model22,model23}. With finite magnetic field, the charged pion condensate breaks both isospin symmetry in the flavor space and translational invariance in the coordinate space, due to its direct interaction with external magnetic field. Furthermore, when one introduces a magnetic field into a pion superfluid, either there is a superconductor or a magnetic vortex, both of which can change the magnetic field. LQCD simulations exhibit a sign problem at finite isospin chemical potential and magnetic field. By using a Taylor expansion in the magnetic field, it is reported that at vanishing temperature, the onset of pion condensate shifts to higher isospin chemical potential under magnetic fields~\cite{lqcdb1}. In the study of pion condensate in effective models, the interaction between the charged pion condensate and the magnetic field is simply neglected in Ref.~\cite{pib1,pib2,pib3}. To avoid this complication, we have investigated the magnetic field effect on pion superfluid phase transition through the Goldstone's theorem~\cite{maopionsuperfluid}, where, starting from the normal phase without pion condensate, the phase boundary of pion superfluid is determined by its massless Goldstone mode ($\pi^+$ meson). Note that the chiral symmetry, which will be restored with increasing isospin chemical potential, controls the mass of quarks, and will influence the formation of pion (quark-anti-quark pair) condensate and the pion superfluid phase transition. However, the previous work on pion superfluid phase transition under magnetic fields do not consider the IMC effect of chiral symmetry restoration.

This paper focuses on the IMC effect on pion superfluid phase transition under magnetic fields. Here we make use of a Pauli-Villars regularized NJL model, which is inspired by the Bardeen-Cooper-Shrieffer (BCS) theory and describes remarkably well the quark pairing mechanisms and the Goldstone mode~\cite{njl1,njl2,njl3,njl4,njl5,zhuang}. Since the interaction between quarks determines the symmetry broken and restoration. In our calculations, the IMC effect is introduced by a magnetic field dependent coupling (see Fig.\ref{geb}). As a straightforward extension of our previous work~\cite{maopionsuperfluid}, we investigate pion superfluid phase transition under magnetic fields through its Goldstone mode ($\pi^+$ meson). %We find that taking into account the IMC effect, no qualitative changes happen and only quantitative differences appear.

The rest of paper is organized as follows. Sec.\ref{sec:f} describes our theoretical framework to study the pion superfluid phase transition including the IMC effect. The numerical results and discussions are presented in Sec.\ref{sec:r}, which focus on the comparison between the results with and without the IMC effect. Finally, we give the summary in Sec.\ref{sec:s}.

\section{Framework}
\label{sec:f}
The two-flavor NJL model is defined through the Lagrangian density in terms of quark fields $\psi$~\cite{njl1,njl2,njl3,njl4,njl5,zhuang}
\begin{equation}
\label{njl}
{\cal L} = \bar{\psi}\left(i\gamma_\nu D^\nu-m_0+\gamma_0 \mu\right)\psi+G \left[\left(\bar\psi\psi\right)^2+\left(\bar\psi i\gamma_5{\vec \tau}\psi\right)^2\right].
\end{equation}
Here the covariant derivative $D_\nu=\partial_\nu+iQ A_\nu$ couples quarks with electric charge $Q=diag (Q_u,Q_d)=diag (2e/3,-e/3)$ to the external magnetic field ${\bf B}=(0, 0, B)$ in $z$-direction through the potential $A_\nu=(0,0,Bx_1,0)$. $m_0$ is the current quark mass. The quark chemical potential $\mu
=diag\left(\mu_u,\mu_d\right)=diag\left(\mu_B/3+\mu_I/2,\mu_B/3-\mu_I/2\right)$ is a matrix in the flavor space, with $\mu_u$ and $\mu_d$
being the $u$- and $d$-quark chemical potentials and $\mu_B$ and
$\mu_I$ being the baryon and isospin chemical potentials. %At finite isospin chemical potential and magnetic field, the isospin symmetry $SU(2)_I$ is broken down to $U(1)_I$ symmetry, and the chiral symmetry $SU(2)_A$ is broken down to $U(1)_A$ symmetry. With the spontaneous breaking of chiral $U(1)_A$ symmetry and isospin $U(1)_I$ symmetry, the Goldstone mode reads $\pi^0$ meson and $\pi^+$ meson, respectively.

$G$ is the coupling constant in scalar and pseudo-scalar channels. In vacuum, the chiral symmetry $U(1)_L \otimes U(1)_R \simeq U(1)_A \otimes U(1)_I$ is spontaneously broken into the isospin symmetry $U(1)_I$. In medium with finite isospin chemical potential, the broken chiral symmetry will be (partially) restored, which leads to the chiral restoration phase transition, and meanwhile, the isospin symmetry will be broken, which leads to the pion superfluid phase transition. Corresponding to the symmetries and their spontaneous breaking, we have two order parameters, neutral chiral condensate $\langle\bar\psi\psi\rangle$ for chiral restoration phase transition and charged pion condensate $\langle\bar\psi\gamma_5\tau^1\psi\rangle$ for pion superfluid phase transition. According to the Goldstone's theorem, the pseudo-Goldstone mode of chiral symmetry breaking is the neutral pion $\pi^0$, and the Goldstone mode of isospin symmetry breaking is the charged pion $\pi^+$. Physically, it is equivalent to define the phase transition by the order parameter and Goldstone mode~\cite{gold1,gold2}.

\begin{figure}[hbt]
\centering
\includegraphics[width=7cm]{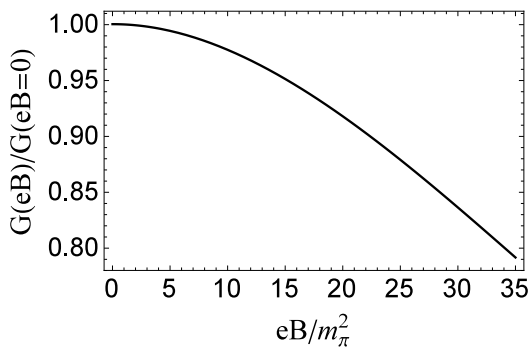}
\caption{Magnetic field dependent coupling $G(eB)$ fitted from LQCD reported decreasing pseudo-critical temperature of chiral symmetry restoration. In this paper, we fix $m_{\pi}=134$ MeV.} \label{geb}
\end{figure}

As a straightforward extension of our previous paper~\cite{maopionsuperfluid}, we use the Goldstone mode (massless $\pi^+$ meson) to determine the pion superfluid phase transition at finite temperature, chemical potential and magnetic field, %In this paper, we mainly focus on the comparison between cases with IMC effect and without IMC effect.
\begin{eqnarray}
m_{\pi^+}(eB,T,\mu_B,\mu_I)=0.
\end{eqnarray}
To include the inverse magnetic catalysis effect, in our calculations, we apply the magnetic field dependent coupling $G(eB)$, which is obtained by fitting the LQCD reported decreasing pseudo-critical temperature of chiral symmetry restoration~\cite{lattice1}. As plotted in Fig.\ref{geb}, the magnetic field dependent coupling $G(eB)/G(eB=0)$ is a decreasing function of magnetic field, and it reduces $20\%$ at $eB/m^2_{\pi} = 35$ (with $m_{\pi}=134$ MeV).

%%%%%%%%%%%%%%%%%%%%%%%%%%%%%%%%%%%%%%%%%%%%%%%%%%%%%%%%%%%%%%%%%%%%
%%%%%%%%%%%%%%%%%%%%%%%%%%%%%%%%%%%%%%%%%%%%%%%%%%%%%%%%%%%%%%%%%%%%

In NJL model, mesons are constructed through quark bubble summations in the frame of random phase approximation~\cite{njl2,njl3,njl4,njl5,zhuang}. Taking into account of the interaction between charged mesons and magnetic fields, the meson propagator $D_{\pi^+}$ can be expressed in terms of the polarization function $\Pi_{\pi^+}$~\cite{maopionsuperfluid,ritus1,ritus2,ritus21,maopimeson},
\begin{eqnarray}
\label{eq4}
D_{\pi^+}({\bar k})=\frac{2G(eB)}{1-2G(eB)\Pi_{\pi^+}({\bar k})}.
\end{eqnarray}
where ${\bar k} =(k_0,0,-\sqrt{(2l+1)eB},k_3)$ is the conserved Ritus momentum of $\pi^+$ meson under magnetic fields.

The meson pole mass $m_{\pi^+}$ is defined through the pole of the propagator at zero momentum $(l=0,\ k_3=0)$,
\begin{eqnarray}
1-2G(eB)\Pi_{\pi^+}(k_0=m_{\pi^+})=0,
\label{pip}
\end{eqnarray}
and
\begin{eqnarray}
\Pi_{\pi^+}(k_0) &=& J_1(m_q)+J_2(k_0),\label{eq7}\\
J_1(m_q) &=& 3\sum_{f,n}\alpha_n \frac{|Q_f B|}{2\pi} \int \frac{d p_3}{2\pi} \frac{1}{ E_f}  \\ &&\times \left[1-f(E_f+\mu_f)-f(E_f-\mu_f) \right],\nonumber \\
J_2(k_0) &=& \sum_{n,n'} \int \frac{d p_3}{2\pi}\frac{j_{n,n'}(k_0)}{4E_n E_{n'}} \label{eq8}\\
&&\times \big[\frac{f(-E_{n'}-\mu_u)- f(E_n-\mu_d)}{k_0+\mu_I+E_{n'}+E_n}\nonumber\\&&+\frac{f(E_{n'}-\mu_u)- f(-E_n-\mu_d)}{k_0+\mu_I-E_{n'}-E_n}\big],\nonumber\\
j_{n,n'}(k_0) &=& \left[{(k_0+\mu_I)^2/2}-n'|Q_u B|-n|Q_d B|\right]j^+_{n,n'} \nonumber\\
&&-2 \sqrt{n'|Q_u B|n|Q_d B|}\ j^-_{n,n'},\label{eq9}
%j_{n,n'}^\pm &=& 2N_c \int dx_1dy_1 {d p_2}/{2\pi} A_\pm(x_1,y_1,p_2),
\end{eqnarray}
with flavors $f=u, d$, spin factor $\alpha_n=2-\delta_{n0}$, quark energy $E_f=\sqrt{p^2_3+2 n |Q_f B|+m_q^2}$, quark Landau level $n=0,1,2...$, (dynamical) quark mass $m_q=m_0-2G(eB)\langle\bar\psi\psi\rangle$ and Fermi-Dirac distribution function $f(x)=1/(e^{x/T}+1)$ in $J_1(m_q)$, and the $u$-quark energy $E_{n'}=\sqrt{p^2_3+2 n' |Q_u B|+m_q^2}$ and $d$-quark energy $E_n=\sqrt{p^2_3+2 n |Q_d B|+m_q^2}$ in $J_2(k_0)$.

The (dynamical) quark mass $m_q$ is determined by the gap equation,
\begin{eqnarray}
1-2G(eB)J_1(m_q)&=&\frac{m_0}{m_q}.
\label{gap}
\end{eqnarray}

Because of the four-fermion interaction, the NJL model is not a renormalizable theory and needs regularization. In this work, we make use of the gauge invariant Pauli-Villars regularization scheme~\cite{njl1,njl2,njl3,njl4,njl5,mao,maopionsuperfluid,maopimeson,rev6}, where the quark momentum runs formally from zero to infinity. The three parameters in the Pauli-Villars regularized NJL model, namely the current quark mass $m_0=5$ MeV, the coupling constant $G(eB=0)=3.44$ GeV$^{-2}$ and the Pauli-Villars mass parameter $\Lambda=1127$ MeV are fixed by fitting the chiral condensate $\langle\bar\psi\psi\rangle=-(250\ \text{MeV})^3$, pion mass $m_\pi=134$ MeV and pion decay constant $f_\pi=93$ MeV in vacuum with $T=\mu_B=\mu_I=0$ and $eB=0$.

\section{results}
\label{sec:r}
%%%%%%%%%%%%%%%%%%%%%%%%%%%%%%%%%%%%%%%%%%%%%%%%%%%%%%%%%%%%%%%%%%%%
\begin{figure}[hbt]
\centering
\includegraphics[width=7cm]{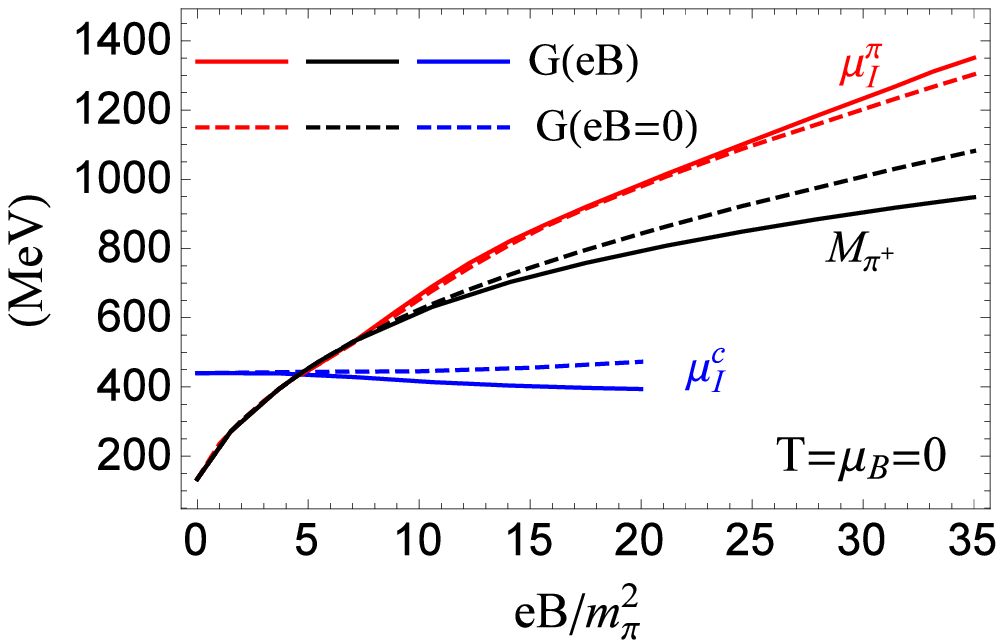}
\includegraphics[width=7cm]{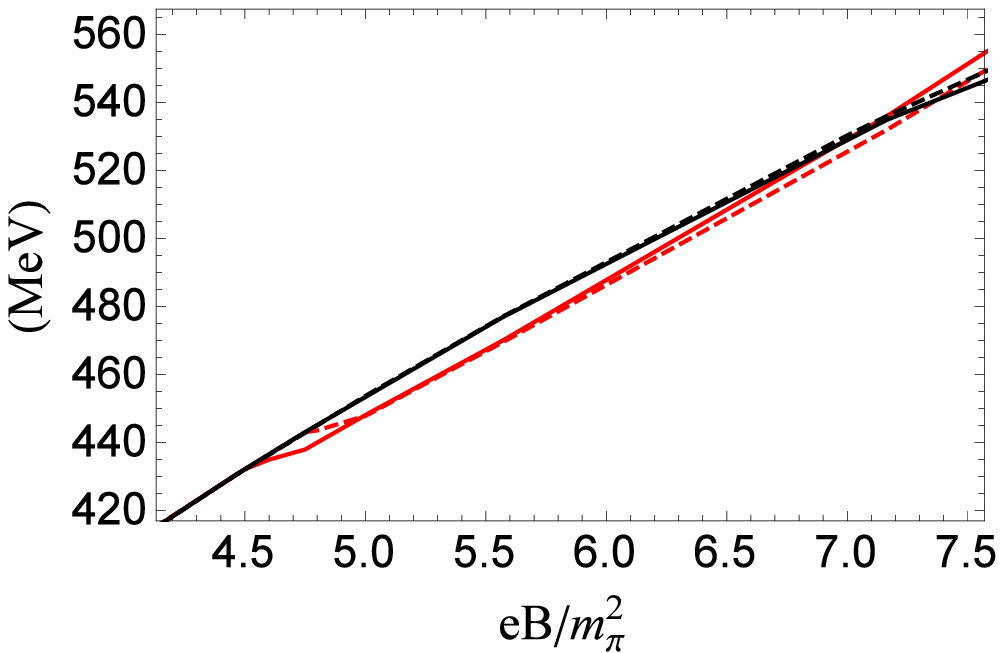}
\caption{(upper panel) Critical isospin chemical potential at $\mu^{\pi}_I$ for pion superfluid phase transition at $T=\mu_B=0$ (red lines), $\pi^+$ meson mass $M_{\pi^+}$ at $T=\mu_B=\mu_I=0$ (black lines) and critical isospin chemical potential $\mu^{c}_I$ for chiral restoration phase transition at $T=\mu_B=0$ (blue lines) as functions of magnetic field with constant coupling $G(eB=0)$ (dashed lines) and magnetic field dependent coupling $G(eB)$ (solid lines). (lower panel) The zoom-in figure in the region $4<eB/m^2_{\pi}<7.5$. In this paper, we fix $m_{\pi}=134$ MeV.} \label{muic}
\end{figure}
%%%%%%%%%%%%%%%%%%%%%%%%%%%%%%%%%%%%%%%%%%%%%%%%%%%%%%%%%%%%%%%%%%%%

Fig.\ref{muic} depicts the critical isospin chemical potential $\mu^{\pi}_I$ for pion superfluid phase transition at $T=\mu_B=0$ (in red), critical isospin chemical potential $\mu^{c}_I$ for chiral symmetry restoration at $T=\mu_B=0$ (in blue) and $\pi^+$ meson mass $M_{\pi^+}$ at $T=\mu_B=\mu_I=0$ (in black) as functions of magnetic field with constant coupling $G(eB=0)$ (dashed lines) and magnetic field dependent coupling $G(eB)$ (solid lines). Here, $\mu^{\pi}_I$ is determined by condition $m_{\pi^+}(eB,T=0,\mu_B=0,\mu_I=\mu^{\pi}_I)=0$, and $\mu^{c}_I$ is defined through the quark mass jump. When including the inverse magnetic catalysis effect by $G(eB)$, the critical isospin chemical potential $\mu^{c}_I$ for chiral symmetry restoration changes from MC phenomenon (increasing with magnetic field) to IMC phenomenon (decreasing with magnetic field). However, the critical isospin chemical potential $\mu^{\pi}_I$ for pion superfluid phase transition with magnetic field dependent coupling $G(eB)$ is similar as the case with constant coupling $G(eB=0)$. It is an increasing function of magnetic field, which means that magnetic field disfavors the pion superfluid phase, even including the IMC effect. There exists some numerical differences. In the regions $eB/m^2_{\pi}<4.5$ and $eB/m^2_{\pi}>5$, the $\mu^{\pi}_I$ with IMC effect (red solid line) is higher than that without IMC effect (red dashed line), and at strong magnetic field, for instance, $eB/m^2_{\pi} = 35$, the difference increases up to $10\%$. Physically, it is harder to form the pion (quark-anti-quark pair) condensate with weaker interaction. Therefore, it is expected to obtain higher $\mu^{\pi}_I$ when including IMC effect. However, with $4.5<eB/m^2_{\pi}<5$, the $\mu^{\pi}_I$ with IMC effect (red solid line) is lower than that without IMC effect (red dashed line). This might be related to the first-order chiral restoration phase transition, associated with an abrupt jump of quark mass, which happens at very similar isospin chemical potential. Note that the crossing point of $\mu_I^c$ and $\mu_I^\pi$ located at $eB/m^2_{\pi}=4.5$ with IMC effect and $eB/m^2_{\pi}=4.75$ without IMC effect, respectively.

In Fig.\ref{muic}, we also make comparison between critical isospin chemical potential $\mu^{\pi}_I$ for pion superfluid phase transition and $\pi^+$ meson mass in vacuum $M_{\pi^+}=m_{\pi^+}(eB,T=0,\mu_B=0,\mu_I=0)$. Under weak magnetic field $eB/m^2_{\pi}<4.5$, the $\mu^{\pi}_I$ is equal to the $\pi^+$ meson mass in vacuum $M_{\pi^+}$, which can be analytically proved by directly comparing the pole equation (\ref{pip}) and gap equation (\ref{gap})~\cite{model9,maopionsuperfluid}. With $4.5<eB/m^2_{\pi}<7$, we obtain $\mu^{\pi}_I < M_{\pi^+}$, and with stronger magnetic field $eB/m^2_{\pi}>7$, we have $\mu^{\pi}_I> M_{\pi^+}$, which are obtained numerically. Without IMC effect, the turning points are located at $eB/m^2_{\pi}=4.75$ and $eB/m^2_{\pi}=7.5$. The deviation between $\mu^{\pi}_I$ and $M_{\pi^+}$ at strong magnetic field is enhanced by the IMC effect.

\begin{figure}[hbt]
\centering
\includegraphics[width=7cm]{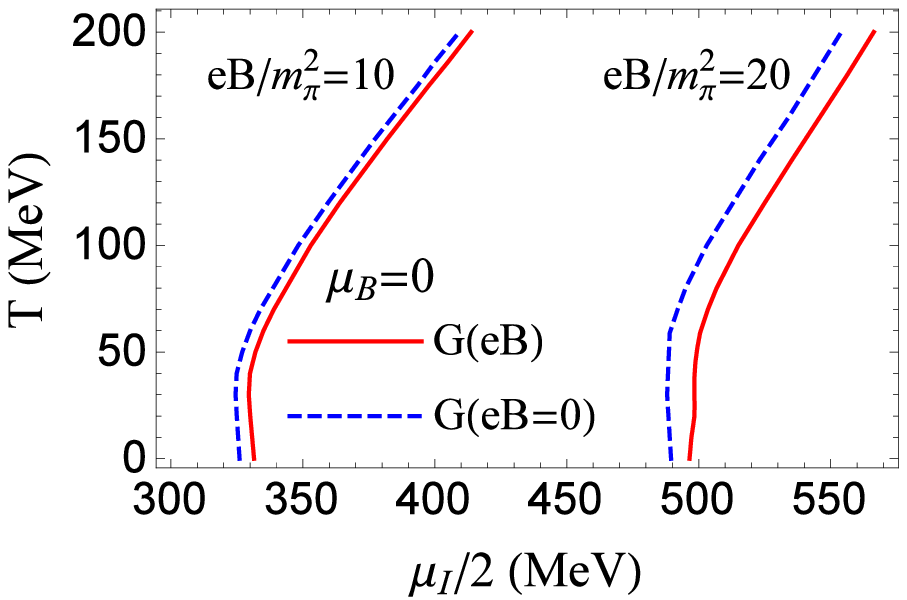}
\includegraphics[width=7cm]{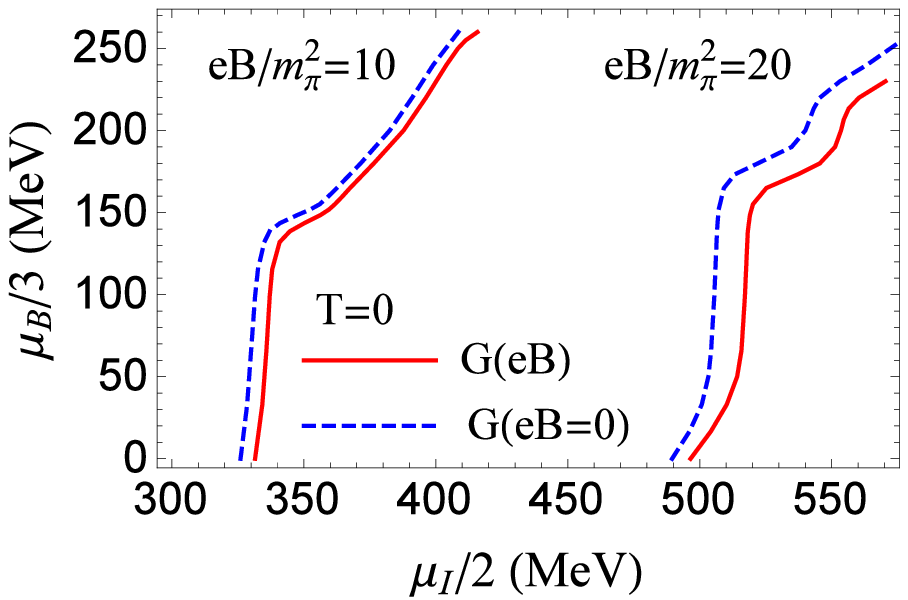}
\caption{(upper panel) Pion superfluid phase diagram in $\mu_I-T$ plane with $\mu_B=0$ and fixed magnetic field. (lower panel) Pion superfluid phase diagram in $\mu_I-\mu_B$ plane with $T=0$ and fixed magnetic field. The left lines are for $eB/m^2_\pi=10$, and right for $eB/m^2_\pi=20$. In this paper, we fix $m_{\pi}=134$ MeV.} \label{tmubmui}
\end{figure}

What is the situation when turning on the temperature and baryon chemical potential? Fig.\ref{tmubmui} is the phase diagram of pion superfluid in $\mu_I-T$ (with $\mu_B=0$) and $\mu_I-\mu_B$ (with $T=0$) planes at $eB/m^2_\pi=10$ and $eB/m^2_\pi=20$, where the solid (dashed) lines correspond to the case with (without) IMC effect. Fixing temperature (upper panel) or baryon chemical potential (lower panel), the phase transition from normal phase to pion superfluid phase happens with increasing isospin chemical potential. On the left side of the phase transition line, it is the normal phase, and on the right side, it is the pion superfluid phase. With higher temperature, the thermal motion of quarks are stronger. Hence it is more difficult to form pion condensate, and the critical isospin chemical potential becomes higher. With fixed temperature and vanishing baryon chemical potential, the critical isospin chemical potential increases with magnetic fields. With higher baryon chemical potential, the mismatch between the Fermi surface of quark and anti-quark is larger. This also prohibits the pion condensate, and leads to higher critical isospin chemical potential. With fixed baryon chemical potential and vanishing temperature, the critical isospin chemical potential also increases with magnetic fields. Including the IMC effect, at finite magnetic field, temperature and baryon chemical potential, the pion superfluid phase transition happens at higher isospin chemical potential, which is caused by the weaker coupling between quark and anti-quark. The difference of critical isospin chemical potential is $\delta \mu^{T}_I \simeq (8\sim 11)$ MeV with fixed $T$ and vanishing $\mu_B$, and $\delta \mu^{\mu_B}_I \simeq (10\sim 19)$ MeV with fixed $\mu_B$ and vanishing $T$ at $eB/m^2_\pi=10$, and $\delta \mu^{T}_I \simeq (14\sim 25)$ MeV and $\delta \mu^{\mu_B}_I \simeq (14\sim 48)$ MeV at $eB/m^2_\pi= 20$. The deviation between the phase transition lines with and without IMC is enhanced by the magnetic field.

%%%%%%%%%%%%%%%%%%%%%%%%%%%%%%%%%%%%%%%%%%%%%%%%%%%%%%%%%%%%%%%%%%%%
%%%%%%%%%%%%%%%%%%%%%%%%%%%%%%%%%%%%%%%%%%%%%%%%%%%%%%%%%%%%%%%%%%%%
\section{summary}
\label{sec:s}
Pion superfluid phase transition under external magnetic field including the inverse magnetic catalysis effect is investigated through the Pauli-Villars regularized NJL model. Based on the Goldstone's theorem, we apply the massless Goldstone boson ($\pi^+$ meson) to determine the phase boundary of pion superfluid. The inverse magnetic catalysis effect is introduced by the magnetic field dependent coupling $G(eB)$, which is a decreasing function of magnetic field.

At fixed temperature and baryon chemical potential, including IMC effect, the critical isospin chemical potential $\mu_I^\pi$ for pion superfluid phase transition increases with the magnetic field, which is qualitatively similar as the case without IMC effect. This indicates that magnetic field disfavors the pion superfluid phase with and without IMC effect. The deviation of $\mu_I^\pi$ with and without IMC effect is enhanced by the magnetic field.

Comparing with the case ignoring IMC effect, the critical isospin chemical potential for pion superfluid phase transition at fixed magnetic field, temperature and baryon chemical potential is shifted to higher value by the IMC effect. There exist different methods to mimic IMC effect in NJL model, which determine the decreasing coupling with magnetic field but with quantitatively different value~\cite{bf8,bf9,geb1,geb2,geb3}. Due to the weakened coupling, it becomes harder to form pion condensate, and the critical isospin chemical potential for pion superfluid phase transition will become higher. This conclusion is independent on the specific formula for coupling $G(eB)$.

%\section{backup}
%Please delete this section.
%
%%%%%%%%%%%%%%%%%%%%%%%%%%%%%%%%%%%%%%%%%%%%%%%%%%%%%%%%%%%%%%%%%%%%%
%\begin{figure}[hbt]
%\centering
%\includegraphics[width=7cm]{mqeb.eps}
%\includegraphics[width=7cm]{mpi+eb.eps}
%\caption{(upper panel) Quark mass as a function of magnetic field in vacuum with constant coupling $G$ and magnetic field dependent coupling $G(eB)$.
%(lower panel) $\pi^+$ meson mass as a function of magnetic field in vacuum with constant coupling $G$ and magnetic field dependent coupling $G(eB)$.} \label{meb}
%\end{figure}
%%%%%%%%%%%%%%%%%%%%%%%%%%%%%%%%%%%%%%%%%%%%%%%%%%%%%%%%%%%%%%%%%%%%%

\noindent {\bf Acknowledgement:}
The work is supported by the NSFC Grant 11775165 and Fundamental Research Funds for the Central Universities.\\

\end{document}